\documentstyle[12pt,psfig,aaspp]{article}
\def\beq{\begin{equation}}
\def\eeq{\end{equation}}

\begin{document}

\title{Does MOND follow from the CDM paradigm?}
\author{Mordehai Milgrom }
\affil{ Institute of Astronomy, Cambridge, UK}

\begin{abstract}

In a recent paper, Kaplinghat and Turner (2001) (KT) advertise
that MOND can be derived naturally in the CDM paradigm. They
actually proceed to produce a more limited result: Every galaxy
should have a transition radius, $r_t$, below which baryons
dominate, and above which dark matter (DM) takes over; the
acceleration at $r_t$ is nearly the same for all galaxies; and due
to a coincidences this is of order $a_0\sim cH_0$. This follows
from their tacit, intermediate result, whereby CDM halos of
galaxies have a very nearly universal acceleration profile
$a(r)\approx v^2(r)/r\approx A\hat a(r/\ell)$, where A is
universal, and only the scale $\ell$ varies from halo to halo.
(This remains so when baryons are added because they assume a
universal baryon-collapse factor.) The KT scenario is
phenomenologically wrong--observed galaxies are simply not like
that. For example, it precludes altogether the existence of LSB
galaxies, in which the acceleration is everywhere smaller than
$a_0$. The phenomenologically sound outcome--i.e., the role of
$a_0$ as a transition acceleration in high-surface-brightness
galaxies--pertains to only a small part of the statement of MOND.
There are several other, independent roles that $a_0\sim cH_0$
plays in MOND phenomenology, and other predictions of MOND, not
related to the value of $a_0$, that are not explainable in the KT
scenario.
 The results of KT also disagree with those of CDM
simulations, which, as they now stand, do not reproduce any aspect
of MOND phenomenology.

\end{abstract}

\section{introduction}
\par
\cite{kt} (KT hereafter) claim in a recent paper that the CDM
paradigm explains why there should appear in the
baryon-dark-matter phenomenology an acceleration $a_0$ that is of
order $cH_0$. They seem to imply that by this they have explained
MOND. In fact, however, they only address one of several
independent roles that $a_0\sim cH_0$ plays in the phenomenology.
They also do not explained predictions of MOND that are not
related to the value of $a_0$. And, furthermore, their argument
leads to predictions that are in clear conflict with observations,
and, in fact, with those of CDM simulations based on the same
physics. Such CDM simulations, which, obviously, are more
reliable, do not reproduce any aspect of the MOND phenomenology;
in particular, they do not point to any acceleration constant of
special significance on galactic scales.

\par
 The KT argument, in paraphrase, is as follows: They
  argue that CDM
 halo formation produces a one-parameter family of halo density
profiles of the form
 \beq \rho(r)=(A/4\pi)S(\ell/\ell_0)G^{-1}\ell^{-1}\hat\rho(r/\ell). \label{density} \eeq
Here, $\ell$ is the present characteristic scale of the halo,
related to the co-moving scale of the perturbation collapsing to
the halo, $L$, by $\ell\propto L/(1+z_c)$, $z_c$ being the
red-shift of non-linearity for the halo. $\ell_0$ corresponds to
$z_c=0$, and $A$ is a constant with the dimensions of
acceleration. Also, one can show that for their halos
 $S(x)= x^{2\theta/(3-\theta)}$, where $\theta=-2-n_{eff}$, with
 $n_{eff}$ the
logarithmic slope of the power spectrum of fluctuations at the
relevant scale. (In KT, this factor appears in terms of $L$ as
$S\propto L^\theta$.)
\par
 Their resulting halo acceleration profile (which KT do not give) is
\beq a(r)=AS(\ell/\ell_0)\hat a(r/\ell), \label{acca} \eeq
 where $\hat
 a(\lambda)\equiv\lambda^{-2}\int_0^\lambda x^2\hat\rho(x)dx$.
 While $n_{eff}$ itself depends on
$\ell$, KT take it very near $-2$, so that $\theta\ll 1$.
Specifically, they take for galaxies $\theta\approx 0.2$, (this is
what they call the first numerical coincidence) giving
$S(x)=x^{0.14}$ for galaxies, which is practically constant over
the range of values of $x$ corresponding to galaxies. Their basic
and main, but unstated, result is then that CDM produces galactic
halos that have a nearly universal acceleration profile, which
differs from halo to halo, practically, only by scaling of the
length.
\par
 Now, continues the argument, the baryon body in all
 galaxies has collapsed by a universal factor of
$\alpha\approx 10$. From this they deduce that all galaxies should
have a transition radius, $r_t=\ell/\alpha$, such that baryons
dominate DM at radii smaller than $r_t$
 and DM takes over at larger radii. The acceleration at that
 radius is [from
eq.(\ref{acca})]  $S(\ell/\ell_0)A\hat a(1/\alpha)$. And, since
$S(\ell/\ell_0)$ hardly varies among galaxies, they get a
universal transition acceleration, which, by another numerical
coincidence, happens to be of the order of $cH_0$.

\section{The KT argument addresses only one of the many roles of $a_0$}
\par
If MOND succeeds in accounting for the mass discrepancy without
dark matter (DM), it tells us that in the DM paradigm, $a_0\sim
cH_0$ appears in several {\it independent} roles in the
phenomenology.

1. In galaxies whose central surface density is high (HSB
galaxies), so that the acceleration in their inner parts becomes
higher than $a_0$, there is a transition radius, $r_t$, which
occurs where the acceleration equal to $a_0$, and which marks the
transition from baryonic to DM dominance. So $a_0$ can be measured
from the transition point.

2. $a_0$ determines the asymptotic acceleration fields in all
galaxies (or, for that matter, in any isolated
  system). Outside the baryonic body, the acceleration at radius
  $r$ is given, asymptotically,
  by $(MGa_0)^{1/2}/r$ where $M$ is the total mass of the galaxy.
  So $a_0$ can be measured from the asymptotics.

  3. $a_0$ defines the transitions from HSBs LSBs:
   Baryons dominate in the inner parts of galaxies whose central
  surface density is higher than some critical value of order
  $a_0G^{-1}$, while in
  galaxies whose central surface density is much smaller,
  DM dominates everywhere.

  4. $a_0$ tells exactly how much DM is needed
  {\it inside the baryon body} in LSBs:
  The lower the mean acceleration, $a$, is, the larger the
  DM-to-baryon
 ratio; specifically, this ratio is predicted to be
 $\approx a_0/a$.

  5. $a_0$ control the dynamics of all galaxy system
such as galaxy groups (\cite{groups}), clusters (\cite{clusters}),
and large-scale filaments (\cite{fil}), where MOND has been shown
to explain away the need for DM. Since MOND has not yet explained
away the need for DM in the inner part of x-ray galaxy clusters,
the DM paradigm is not yet called upon to explain such a success.
Clusters at large, however--say within two megaparsecs of the
center are explained by MOND (\cite{clusters}).

Strictly speaking, no two galaxies or galactic systems have had
exactly the same history of formation-evolution-interaction.
Inasmuch as MOND succeeds in obtaining the DM distribution from
the baryon mass distribution in any such system, with only the aid
of $a_0$--as it claims to be capable of--each such success would
call for a separate explanation in the DM paradigm. The DM
paradigm has, at best, tried to address general trends in the
population behavior of galactic systems. This, indeed, is the most
that can be asked from it, in light of the expected idiosyncrasies
of individual objects. But why then should there be a simple
theory that does account for these idiosyncrasies?

These appearances of $a_0$ in the DM phenomenology are independent
because a-priori one could (easily) envisage baryon-plus-DM
galaxies and galactic systems in which $a_0$ appears in any of the
above roles but not in the others. For example, if halos are cut
off beyond some radius larger than $r_t$ we can have property 1
but not 2 (or 3 if the cutoff radius is small compared with the
inter-galaxy distance). It follow, for example, that the
appearance of $a_0$ in the TF relation is independent of its role
as a transition acceleration. Another example, galaxies might have
been such that they do not become more and more DM dominated as
their central surface density decreases; so property 4 will not
have been satisfied, but all the other MOND predictions could
still be retained.
\par
The KT galaxies are yet another example. They exhibit  appearance
1 of $a_0$, but not the others. To explain 2 one would have to
assume an asymptotic $r^{-2}$ halo density profile, which is not a
consequence of CDM. Appearances 3 and 4 in the phenomenology
actually fly in the face of the KT argument because
 all their galaxies have the same surface density (see details in
 the next
section). The KT argument is said to apply on the scale of
galaxies only, so does not pertain to appearance 5 of $a_0$. In
fact, at larger scales $n_{eff}\rightarrow -3$ so
$\theta\rightarrow 1$, and $S(\ell/\ell_0)$ becomes linear in
$\ell$. By the KT picture we would then expect MOND phenomenology
to miss by orders of magnitudes on large scales (hundreds of times
larger than galactic scales). In fact, this is not so. And, even
in galaxy clusters (say within 1-2 Mpc) where MOND does not yet
explain away all the DM, the remaining gap is only of a factor of
order 2, much smaller than what the KT scenario predicts.
\par
We learn from all this that if one purports to explain MOND and
the appearance of $cH_0$ in the phenomenology, within the DM
paradigm, he should explain all the above appearances, because
they do not follow from each other within this paradigm. One of
the strengths of MOND is that it does tightly connect all those
apparently different occurrences (MOND was constructed to account
for RC asymptotics and all the others came as unavoidable
predictions).
  In the DM paradigm this is still an unexplained miracle.

\section{The KT predictions conflict with observations}

\par
 An unavoidable prediction of the
 KT argument is that {\it all} galaxies must have a
transition radius--occurring at a fixed fraction of the halo's
length scale--where the halo acceleration goes from $a>a_0$ to
$a<a_0$. This follows from eq.(\ref{acca}), which implies that all
halos have very nearly the same acceleration runs, so either no
galactic halo has reaches an acceleration value of $a_0$ or they
all do. In particular, there should be a range of radii, at least
up to $r_t$, where the halo acceleration is above $a$ (adding the
baryons changes the acceleration profile but not this fact).
 This (which is not
a prediction of MOND) blatantly conflicts with observations:
 Many galaxies have
 accelerations--measured directly from their rotation curves--that
  are everywhere much smaller than $a_0$
  (generally classified as LSB galaxies). According to KT such galaxies
 are not produced in CDM.
 \par
This problem is yet augmented by the observation of pairs of
galaxies whereby the two galaxies in the pair have  the same total
luminosity
  and the same asymptotic rotational
 velocity--so they lie on the same point of the TF relation--but
 very different acceleration profiles.
  In particular, there are such pairs with one galaxy being an HSB and
  the other an LSB.
According to the KT scenario, the two galaxies in such a pair have
the same halo,
  and, in particular, should have the same transition radius and
  acceleration.
 Yet, this is clearly not so. This point is demonstrated by \cite{bm}
  with the pair NGC 2403 and UGC 128.
The former is an HSB
 having a clear transition from a super-$a_0$ region near the
 center to the sub-$a_0$ region, occurring at about 10 kpc--several
 disc scale lengths, while the latter is an LSB with accelerations
  everywhere
  smaller that $a_0$. \cite{tv} have further emphasized this
  point, showing that such pairs are quite common.
  \par
  In the KT picture, baryons in all galaxies had undergone
  a collapse by the same factor. The resulting baryon-plus-DM
  systems all have the same surface-density distribution, again,
  differing only by their scale length. In particular, they all should
   have the same rotation curve. This is obviously not true
   observationally.
\par
It need be emphasized, perhaps, that those features of the KT
scenario that lead directly to clashes with observations--the near
universality of the halo surface density, and of the baryon
collapse--are indispensable; without them the argument itself
collapses, and no universal transition acceleration emerges.

\section{Is the argument valid at all?}
\par
Do the halos that KT required for their argumentation actually
follow in the CDM paradigm? The best way to answer this, at
present, is to consult the results of numerical CDM simulations,
which start from the same physics. The fact is that such
simulations do give approximately a one-parameter halo family, but
not of the scaling deduced by KT--see e.g. \cite{nfw}. In
particular, no constant with the dimensions of acceleration
emerges from  the simulations on galactic scales,as it does in KT.
This disparity is evident in the fact that the KT halos have a
mass velocity relation $M\propto v^{3.75}$, while the simulations
of \cite{nfw}, for example, give $M\propto v^\delta$ with
$\delta\approx 3.2-3.3$. A power of 3.3 corresponds to
$S(x)=\approx x^{0.5}$, which gives a large variation of the
transition acceleration among galaxies.
\par
Perhaps this disparity can be traced back to the value of
$n_{eff}$ that KT chose to suit their purpose, but which is not
what the simulations dictate; perhaps it is due to details of the
more exact simulations that are not captured by the argumentation
of KT. If it is the former reason, we can note that to obtain an
$M-v$ power of 3 one needs $\theta=1$, and to get a power of 3.3
one needs $\theta=0.66$, compared with the rather smaller value of
$0.2$ that KT took.
\par
Beyond all this, there remains the fact that the KT scenario is
very crude. It assumes a universal collapse factor for the
baryons; it does not include the contribution of the baryons
themselves to the acceleration; it does not allow for baryon
accretion or expulsion; and, it neglects the response of the halo
to baryon collapse. Moreover, the KT result depends crucially on
their assumption that a large fraction of the original baryons in
the halo had collapsed to form the observed galaxy. In contrast,
it can be argued quite cogently that only a small fraction of the
original baryons go into forming the observed galaxy. This follows
from the high values of DM-to-observed-baryons mass ratios deduced
for galaxies--e.g. in \cite{bm}, \cite{zar}, and \cite{mk}--which
are much higher than the cosmological value of this parameter,
believed now to be of order 7-10.


\end{document}